\documentclass[a4paper,11pt]{article}
\usepackage{jcappub} % for details on the use of the package, please see the JINST-author-manual
\usepackage{lineno}
\usepackage{amsmath}
\usepackage{amsfonts}
\usepackage{amssymb}
\usepackage{mathrsfs}
\usepackage{graphicx}
\usepackage{color}
\usepackage[utf8]{inputenc}
\usepackage{amsfonts}
\usepackage{graphicx}
\usepackage{booktabs}
\usepackage{multirow}
\usepackage{makecell}
\usepackage{ragged2e}
\usepackage{rotating}
\usepackage{hyperref}
\usepackage{float}
\usepackage{booktabs}
\usepackage{xcolor}  
\usepackage{orcidlink}
\definecolor{darkred}{rgb}{0.5,0,0}
\definecolor{darkblue}{rgb}{0,0,0.5}
\definecolor{firebrick}{rgb}{0.75,0.125,0.125}
\definecolor{darkgreen}{rgb}{0,0.5,0}

\hypersetup{
    colorlinks,
    linkcolor={cyan!80!black},
    citecolor={cyan!80!black},
    urlcolor={cyan!80!black}
}        
\usepackage{lipsum}
\usepackage[normalem]{ulem}
\usepackage{enumitem}
\usepackage[capitalise]{cleveref}
\usepackage{cleveref}
\usepackage{subfigure}
\usepackage{physics}
\usepackage{comment}
\usepackage{tabularx}  
\newcommand{\sixflav}[6]{(\{{#1},{#2}\}:\{#3,#4\}:\{#5,#6\})}
\newcommand{\sixflavor}[6]{\{{#1},{#2}\}:\{#3,#4\}:\{#5,#6\}}
\newcolumntype{Y}{>{\centering\arraybackslash}X}
\newcolumntype{Z}[1]{>{\centering\arraybackslash}m{#1}}

\graphicspath{{figs/}}

\title{Probing Neutrino Flavor Composition with the Glashow Resonance at Tau Air-Shower Neutrino Telescopes }

\author[a,b]{Tianyi Ding\,\orcidlink{0009-0004-4928-2763}}
\author[c,d]{Qinrui Liu\,\orcidlink{0000-0003-3379-6423}}

\affiliation[a]{Dept. of Physics \& Astronomy, University of Nevada Las Vegas, NV, 89154}
\affiliation[b]{Nevada Center for Astrophysics (NCfA), University of Nevada Las Vegas, NV, 89154}
\affiliation[c]{
Department of Physics, Simon Fraser University, Burnaby BC V5A 1S6, Canada
}
\affiliation[d]{ Arthur B. McDonald Canadian Astroparticle Physics Research Institute, Kingston ON K7L 3N6, Canada}

\emailAdd{dingt2@unlv.nevada.edu}
\emailAdd{qinrui\_liu@sfu.ca}

\abstract{The flavor composition of high-energy astrophysical neutrinos encodes information about their production and propagation. The Glashow resonance, $\bar{\nu}_e + e^-\to W^-$, provides a unique way to distinguish antineutrinos from neutrinos and thereby extends the reach of flavor composition studies. Proposed tau air-shower neutrino telescopes target Earth-skimming and mountain-skimming $\nu_\tau$ above a PeV, but through the decay $W^-\to \bar{\nu}_\tau+\tau^-$ they are also sensitive to $\bar{\nu}_e$. These experiments can therefore measure the ratio of $\bar{\nu}_e$ to $\nu_\tau+\bar{\nu}_\tau$ fluxes. We evaluate this prospect with explicit simulations and project sensitivities for TAMBO and TRINITY, assuming 10 years of operation. We find that mountain-skimming geometries yield substantially higher $\bar{\nu}_e$ acceptance than Earth-skimming ones due to the shorter path length in rock. For standard astrophysical source scenarios, our projections show that differentiating $pp$ and $p\gamma$ production, including their muon-damped scenarios, is challenging with a standalone measurement by tau air-shower experiments in their currently designed configurations, though optimistically the flux ratio can be constrained to $\lesssim$2 at 1$\sigma$. A $\bar{\nu}_e$-rich flux, as expected from neutron-decay sources or from certain new physics models, would stand out from the standard pion-production scenarios and can otherwise be constrained.
}

\begin{document}
\maketitle

\section{Introduction}
\label{sec:intro}

The detection of TeV--PeV neutrinos by IceCube opened a new window to the high-energy universe~\cite{IceCube:2013low}, and neutrino astronomy has been flourishing in recent years, including significant insights into the origin of the observed diffuse neutrino flux with the identification of neutrino emission from the Milky Way and active galactic nuclei~\cite{IceCube:2018dnn,IceCube:2018cha,IceCube:2022der,IceCube:2023ame}. 

Neutrinos carry flavors, which provides us with a unique venue to investigate the neutrino production mechanisms at the astrophysical sources, neutrino propagation effects to the Earth, and physics in the detector by studying the flavor composition of the neutrino flux observed in the detector.  
Extensive work has examined the flavor composition of high-energy neutrinos observed~\cite{Mena:2014sja,Palomares-ruiz:2015mka,IceCube:2015rro,IceCube:2015gsk,Vincent:2016nut,Song:2020nfh,IceCube:2020fpi,Abbasi:2025fjc} and discussed its implications for astrophysics and neutrino physics, especially new physics, \textit{e.g.} in Refs.~\cite{Learned:1994wg,Athar:2000yw,Crocker:2001zs,Beacom:2002vi,Barenboim:2003jm,Beacom:2003nh,Beacom:2003eu,Beacom:2003zg,Serpico:2005bs,Mena:2006eq,Kachelriess:2006fi,Pakvasa:2007dc,Esmaili:2009dz,Choubey:2009jq,Esmaili:2009fk,Bhattacharya:2009tx,Bhattacharya:2010xj,Bustamante:2010nq,Mehta:2011qb,Baerwald:2012kc,Fu:2012zr,Pakvasa:2012db,Chatterjee:2013tza,Xu:2014via,Aeikens:2014yga,Palladino:2015zua,Arguelles:2015dca,Bustamante:2015waa,Pagliaroli:2015rca,deSalas:2016svi,Gonzalez-Garcia:2016gpq,Bustamante:2016ciw,Rasmussen:2017ert,Dey:2017ede,Bustamante:2018mzu,Farzan:2018pnk,Ahlers:2018yom,Brdar:2018tce,Palladino:2019pid,Bustamante:2019sdb,Ahlers:2020miq,Karmakar:2020yzn,Fiorillo:2020gsb,Dev:2024yrg,Abbasi:2025fjc,Bustamante:2026aur,Bustamante:2026zst}. While most studies so far focus on the flavor composition averaged over a wide energy range, the flavor composition might be energy dependent due to astrophysical effects such as the dominant neutrino production mechanism changes with energy~\cite{Kashti:2005qa, Lipari:2007su, Hummer:2010ai,Baerwald:2011ee, Bustamante:2020bxp, Fiorillo:2021hty, Bhattacharya:2023mmp} or neutrino physics where various non-standard effects can affect the oscillations of neutrinos in an energy-dependent manner, see {\textit{e.g.}} Refs.~\cite{Minakata:1996nd,Beacom:2002vi,Beacom:2003eu,Mehta:2011qb,Bustamante:2015waa,Arguelles:2015dca,Shoemaker:2015qul,Karmakar:2020yzn,Abdullahi:2020rge}. Ref.~\cite{Liu:2023flr} studies the identification of energy-dependent flavor transitions in TeV-PeV optical Cherenkov detectors. Such measurements are expected to probe astrophysical processes and to test neutrino physics independently at higher energies, complementing high-energy measurements and enabling the examination of energy dependence in the flavor composition should energy-dependent transitions occur beyond the reach of optical telescopes.   

At the same time, the majority of these studies have focused on the composition of 3 neutrino flavors, assuming that there is an equal contribution from neutrinos and antineutrinos, {\textit{i.e.}} when discussing the neutrino flavor composition, neutrinos and antineutrinos are usually not separated. This is because the neutrino-nucleon deep inelastic scattering (DIS) cross section for neutrinos and antineutrinos coincide with each other as the neutrino energy goes higher, especially above $\mathcal{O}\left(100\rm{TeV}\right)$, making it impossible for an all-flavor neutrino telescope such as IceCube to distinguish neutrinos and antineutrinos~\cite{Gandhi:1995tf,Formaggio:2013kya}. However, due to the abundant electrons on Earth, the resonant scattering around the $W^-$ pole~\cite{Glashow:1960zz}, {\textit{i.e.}} the Glashow resonance $\bar{\nu}_e + e^-\rightarrow W^-$, significantly enhances the cross section around the neutrino energy of 6.3~PeV, which provides an effective way to differentiate the $\bar{\nu}_e$ flux from the total flux. Such an event was first observed by IceCube~\cite{IceCube:2021rpz}. The detection of Glashow resonant events with neutrino telescopes can provide deeper insight into astrophysical processes~\cite{Brown:1981ns,Anchordoqui:2004eb,Bhattacharya:2011qu,Xing:2011zm,Barger:2012mz,Bhattacharya:2012fh,Anchordoqui:2014yva,Barger:2014iua,Palladino:2015uoa,Anchordoqui:2016ewn,Biehl:2016psj,Huang:2019hgs,Huang:2023yqz,Liu:2023lxz,Huang:2023mgt} and open new avenues for new physics such as neutrino decay, Lorentz invariance violation and asymmetric dark matter~\cite{Stecker:2014oxa,Shoemaker:2015qul,Bustamante:2020niz,Xu:2022svm,Liu:2024wmk}.

In the upcoming decades, new detectors are being planned to target the neutrino energy range above PeV, where IceCube becomes less sensitive to detecting a significant flux, and to do so via various approaches. Tau air-shower neutrino telescopes such as Ashra Neutrino Telescope Array (Ashra NTA), Tau Air
Shower Mountain-Based Observatory (TAMBO) and TRINITY~\cite{Sasaki:2014mwa,Romero-Wolf:2020pzh,TAMBO:2025jio,Otte:2019knb} are designed to detect neutrinos with a threshold $\sim$1~PeV, while there are multiple other experiments focusing on ultra-high-energy neutrino detection with a threshold above tens of PeV. For such a telescope, only tau neutrinos are expected to be observed, because unlike taus, electrons and muons do not produce extensive air showers. However, as there is an 11\% branching ratio of the decay mode $W^-\rightarrow \tau^- + \bar{\nu}_\tau$ and the Glashow resonant cross section is about 2 orders of magnitude larger compared to the DIS cross section, we should expect the tau flux at PeV contains a non-negligible contribution from $\bar{\nu}_e$ in addition to tau neutrinos. While the potential detection of Glashow resonant induced tau events is discussed in several previous works~\cite{Huang:2019hgs,Liu:2023lxz,Huang:2023mgt}, the prospects for a flavor measurement and the comparison of different detector geometries remain largely unexplored. The $\bar{\nu}_e$ contribution opens up the possibility of standalone studies of the neutrino flavor composition within tau air-shower telescopes, by statistically measuring the relative contributions of $\bar{\nu}_e$ and $\nu_\tau+\bar{\nu}_\tau$. No combination with telescopes optimized for higher or lower energies is required. This also provides an exclusive measurement of the flavor composition at PeV-100~PeV energies alone, which fills the gap between the observations of high-energy and ultra-high-energy neutrino telescopes,  which is crucial for future attempts to resolve the energy-dependence of the flavor composition by comparing the measurements at different energy ranges. 

In this work, we run detailed simulations to evaluate the propagation of neutrinos and the detection of tau air showers induced by both DIS of tau neutrinos and Glashow resonance for two cases where neutrinos interact inside mountains (mountain-skimming) or near the surface below the horizon (Earth-skimming) in proposed TAMBO~\cite{Romero-Wolf:2020pzh,TAMBO:2025jio} and TRINITY~\cite{Otte:2019knb} neutrino telescopes which aim for optimal neutrino detection sensitivities at very-high-energies (PeV-EeV). Then we project their capability of differentiating $\nu_\tau/\bar{\nu}_\tau$ and $\bar{\nu}_e$ and evaluate the effect of the additional flux from Glashow resonance in the spectrum measurement. We do not discuss other ultra-high-energy neutrino telescopes in this work, as their detection thresholds are usually above tens of PeV, where no Glashow Resonant events are expected. 

The rest of the paper is structured as follows. In Sec.~\ref{sec:signal}, we compute $\tau$ signal from $\bar{\nu}_e$ and $\nu_\tau+\bar{\nu}_\tau$ for both the mountain-skimming and Earth-skimming cases from simulations. In Sec.~\ref{sec:flavor}, we present the sensitivity of measuring the flavor ratio for multiple scenarios. We conclude in Sec.~\ref{sec:conclusion}.  

\section{ \texorpdfstring{$\tau$}{tau} signal from \texorpdfstring{$\bar{\nu}_e$}{nuebar} via Glashow Resonance}
\label{sec:signal}

Tau air-shower experiments aim to observe the extensive air shower induced by the decay of high-energy tau leptons. These taus are produced via charged-current interactions of tau neutrinos in the Earth or mountains and must escape the matter surface before decaying, with a decay length $c\tau_\tau \gamma \simeq 50\,{\rm{m}}\left(E_\tau/\rm{PeV}\right)$. The resulting showers are observed through techniques such as Cherenkov light, fluorescence, and radio emission. Although the Earth becomes opaque to neutrinos at high energies as the cross section grows, tau regeneration~\cite{Halzen:1998be,Learned:1994wg,Beacom:2001xn,Dutta:2002zc,Bugaev:2003sw} enables detection of the tau signal after traversing a long distance in matter: high-energy taus produced in charged-current $\nu_\tau$ interactions decay back into neutrinos, and the resulting $\nu_\tau \leftrightarrow \tau$ conversion chain yields an observable flux at lower energies. Tau leptons and tau neutrinos produced in $W^-$ decays introduce an additional contribution, originating from $\bar{\nu}_e$. This contribution cannot be separated from the primary astrophysical $\nu_\tau+\bar{\nu}_\tau$ flux on an event-by-event basis, but it can be studied statistically. 

The expected total tau events can be estimated by computing the propagation of neutrinos and taus. When accounting for the Glashow resonant interactions, $\bar{\nu}_e$ propagation can be written as

\begin{equation}\label{eq:nue}
\begin{aligned}
\frac{d^2 \Phi_{\bar{\nu}_e}}{d x d E_\nu}= & -\sum_{N=p,n} n_N(x)\left[\sigma_{\mathrm{CC}}(E_\nu)+\sigma_{\mathrm{NC}}(E_\nu)\right] \frac{d\Phi_{\bar{\nu}_e}}{dE_\nu} -  n_e(x)\sigma_{\rm{GR}}(E_\nu)\frac{d\Phi_{\bar{\nu}_e}}{dE_\nu} \\ & +\sum_{N=p,n} n_N(x) \int_{E_\nu} d E_\nu^{\prime} \frac{d\Phi_{\bar{\nu}_e}}{d E_\nu^{\prime}} \frac{d\sigma_{\mathrm{NC}}}{d E_\nu}\left(E_\nu^{\prime}, E_\nu \right) +\int_{E_\nu} d E_\tau^{\prime} \frac{d \Phi_{\tau^{\pm}}}{d E_{\tau}^{\prime}} \frac{1}{E_\tau^{\prime}} \frac{d N_{\tau^-\rightarrow {\bar{\nu}_e}}}{dy} \\&
+ n_e(x) \int_{E_\nu} dE_\nu^{\prime} \frac{d \Phi_{\bar{\nu}_e}}{d E_\nu^{\prime}} \sigma_{\mathrm{GR}}\left(E_\nu^{\prime}\right)\frac{1}{E_\nu^{\prime}} \frac{d N_{W^- \rightarrow \bar{\nu}_e}}{d y},
\end{aligned}
\end{equation}
where $\sigma_{\mathrm{CC}}$ and $\sigma_{\mathrm{NC}}$ are charged-current and neutral-current DIS cross sections between neutrinos and nucleons, while $\sigma_{\rm{GR}}$ is the Glashow resonant cross section. Here, we denote $\Phi_{\tau^{\pm}} = \Phi_{\tau^+}+\Phi_{\tau^-}$. The regeneration term implicitly selects the $\tau^-$ component when computing $\bar{\nu}_e$ production. The densities of proton, neutron, and electrons have $n_p = n_e = \frac{Z}{A}N_A\rho$ and $n_n = \frac{A-Z}{A}N_A\rho$, where $Z$ is the atomic number, $A$ is the mass number, $N_A$ is Avogadro's constant, and $\rho$ is the mass density at location $x$, assuming isoscalar matter. The last three terms represent the regeneration of $\bar{\nu}_e$ from the neutral-current interactions of $\bar{\nu}_e$ at higher energies, as well as the decay of the $\tau^-$ flux and $W^-$ produced by the Glashow resonant interactions. Here, $d N_{W^- \rightarrow i}/d y$ and $d N_{\tau\rightarrow i}/dy$ represent the decay spectra of $W^-$ and tau leptons to particle $i$ while $y$ is the energy fraction between $i$ and the parent particle. 

Due to the 11\% tau decay mode of $W^-$, the propagation of $\nu_\tau+\bar{\nu_\tau}$ and $\tau^\pm$ can be updated. The $\nu_\tau+\bar{\nu_\tau}$ flux can similarly be written as 

\begin{equation}\label{eq:nutau}
\begin{aligned}
\frac{d^2 \Phi_{\nu_\tau+\bar{\nu}_\tau}}{d x d E_\nu}= & -\sum_{N=p,n} n_N(x)\left[\sigma_{\mathrm{CC}}(E_\nu)+\sigma_{\mathrm{NC}}(E_\nu)\right] \frac{d\Phi_{\nu_\tau+\bar{\nu}_\tau}}{d E_\nu} \\ &+\sum_{N=p,n} n_N(x) \int_{E_\nu} d E_\nu^{\prime} \frac{d\Phi_{\nu_\tau+\bar{\nu}_\tau}}{d E_\nu^{\prime}} \frac{d\sigma_{\mathrm{NC}}}{d E_\nu}\left(E_\nu^{\prime}, E_\nu \right)  \\
& +\int_{E_\nu} d E_\tau^{\prime} \frac{d \Phi_{\tau^\pm}}{d E_\tau^{\prime}} \frac{1}{E_\tau^{\prime}} \frac{dN_{\tau\rightarrow {\nu_\tau/\bar{\nu}_\tau}}}{dy} +n_e(x)  \int_{E_\nu} dE_\nu^{\prime} \frac{d \Phi_{\bar{\nu}_e}}{d E_\nu^{\prime}} \sigma_{\mathrm{GR}}\left(E_\nu^{\prime}\right)\frac{1}{E_\nu^{\prime}} \frac{d N_{W^- \rightarrow \nu_\tau/\bar{\nu}_\tau}}{dy},
\end{aligned}
\end{equation}
where $\nu_\tau+\bar{\nu_\tau}$ from the $W^-$ decay emerges in the tau neutrino flux. In addition to $W^-\rightarrow \tau^-+ \bar{\nu}_\tau$, there is a flux of $\tau^+$ and $\nu_\tau$ at a lower level from heavy flavor hadrons in the hadronic shower of $W^-$ decay. The propagation of $\nu_e$, $\nu_\mu$ and  $\bar{\nu}_\mu$ can be written similarly to Eq.~\ref{eq:nutau} as they can be produced via $\tau$ and $W^-$ decays. However, they are not expected to contribute to the final detectable signal in a tau air-shower detector. 

The propagation of $\tau$ can be written as 
\begin{equation}\label{eq:tau}
\begin{aligned}
\frac{d^2 \Phi_{\tau^\pm}}{d x d E_\tau}= & -\Gamma_\tau \frac{d\Phi_{\tau^\pm}}{dE_\tau} - g(E_\tau, x) \frac{d\Phi_{\tau^\pm}}{dE_\tau}  + \int_{E_\tau} f(E_\tau^\prime,E_\tau, x)\frac{d\Phi_{\tau^\pm}}{dE_\tau^\prime}dE_\tau^\prime \\ & +\sum_{N=p,n} n_N(x) \int_{E_\tau} d E_\nu^{\prime} \frac{d\Phi_{\nu_\tau+\bar{\nu}_\tau}}{d E_\nu^{\prime}} \frac{d\sigma_{\mathrm{CC}}}{d E_\tau}\left(E_\nu^{\prime}, E_\tau \right) 
\\ & +n_e(x) \int_{E_\tau} dE_\nu^{\prime} \frac{d \Phi_{\bar{\nu}_e}}{d E_\nu^{\prime}} \sigma_{\mathrm{GR}}\left(E_\nu^{\prime}\right)\frac{1}{E_\nu^{\prime}} \frac{d N_{W^- \rightarrow \tau}}{ d y},
\end{aligned}
\end{equation}
where $\Gamma_\tau$ is the decay width of $\tau$ and $g(E_\tau,x)$ describes the energy loss of $\tau$ at $x$, which is mainly through ionization, bremsstrahlung, pair production, and photo-nuclear interactions~\cite{koehne2013proposal}. Meanwhile, $\tau$ at a higher energy loses energy to $E_\tau$ and can be described by a function $f(E_\tau^\prime,E_\tau, x)$. For the Glashow resonant cross section $\sigma_{\rm{GR}}$, in addition to the resonant cross section, we also include corrections from the Doppler broadening~\cite{Loewy:2014zva} and the initial photon radiation~\cite{Gauld:2019pgt,Garcia:2020jwr}. These effects lead to a $\sim$30\% lower value at the resonant energy, a slight broadening, and a larger cross section above the resonant energy~\cite{Huang:2023yqz}. Here, we ignore the
subdominant scattering processes such as the trident production~\cite{Seckel:1997kk,Alikhanov:2015kla,Zhou:2019vxt,Zhou:2019frk}.

\subsection{Simulation}
\label{Simulation_sec}
The propagation of tau leptons and neutrinos is implemented in various neutrino propagation codes such as \texttt{nuFATE}~\cite{Vincent:2017svp}, \texttt{NuTauSim}~\cite{Alvarez-Muniz:2018owm}, \texttt{NuPropEarth}~\cite{Garcia:2020jwr}, \texttt{nuPyProp}~\cite{Garg:2022ugd} and \texttt{TauRunner}~\cite{Safa:2019ege,Safa:2021ghs}. In this work, we employ a modified \texttt{TauRunner} by adding the Glashow resonant interactions to simulate the propagation of neutrinos and tau leptons in Earth or mountains, and obtain the tau flux $\Phi_{\tau^\pm}$ entering the air which is generated from $\nu_\tau + \bar{\nu}_\tau+{\bar{\nu}_e}$. In the public \texttt{TauRunner} framework, neutrino propagation includes neutrino--nucleon DIS and the associated $\tau$ production, propagation, decay, and regeneration, but does not include neutrino--electron interactions. We modify \texttt{TauRunner} by explicitly including the Glashow resonant interaction, treated consistently alongside neutrino--nucleon DIS, according to Eqs.~\ref{eq:nue}--\ref{eq:tau}. The spectra of decay products of $W^-$ are generated using \texttt{PYTHIA8.3}~\cite{Bierlich:2022pfr}. All secondary leptons and neutrinos are subsequently propagated using the existing \texttt{TauRunner} machinery. 

\begin{figure}[t]
    \centering
    \includegraphics[width=1\textwidth]{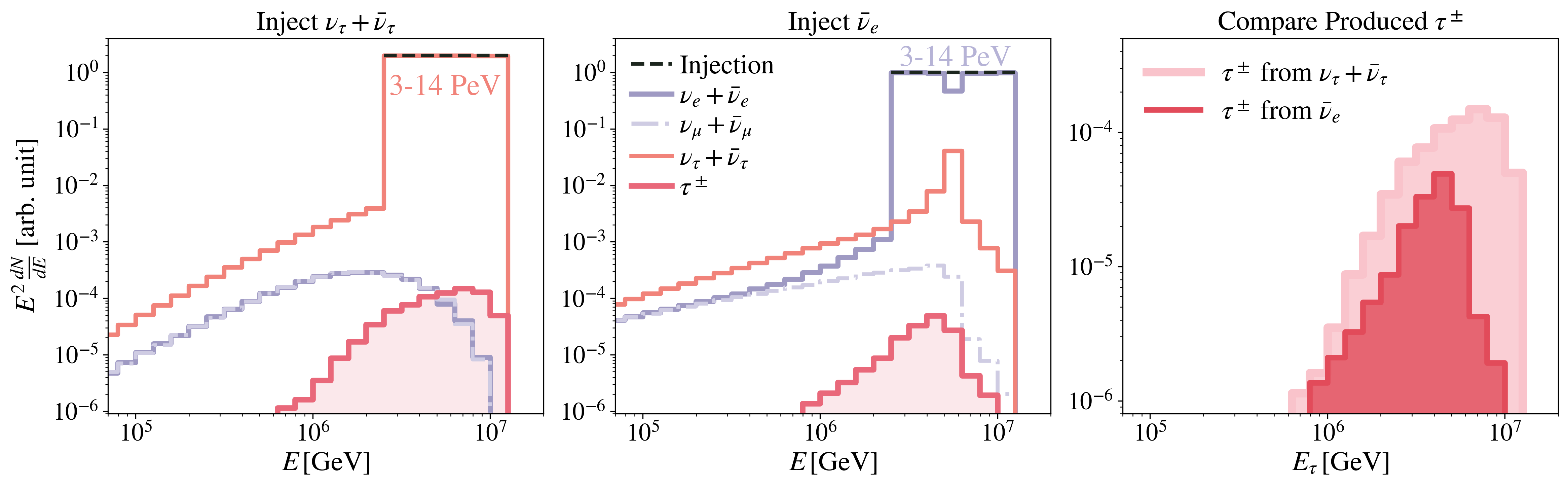}
    \caption{{\bf{\textit{Energy Spectra at Surface for mountain-skimming neutrinos.}}} Energy spectra of particles emerging after neutrino propagation through 12~km of standard rock. The neutrino injection of a flat spectrum $E^{-2}$ covers an energy range 3-14~PeV. The left panel is the simulated result of injecting $\nu_\tau+\bar{\nu}_\tau$, while the middle panel is that from injecting $\bar{\nu}_e$. The injected $\nu_\tau+\bar{\nu}_\tau$ flux is normalized to twice the injected $\bar{\nu}_e$ flux, corresponding to an equal flavor composition benchmark, where the black dashed lines indicate the injection level. The fluxes of neutrinos of all flavors and the final tau leptons (red) are displayed. The right panel shows the tau fluxes generated from the $\bar{\nu}_e$ and $\nu_\tau+\bar{\nu}_\tau$ fluxes for comparison.}
\label{fig:sim_result}
\end{figure}

For the mountain-skimming case, Fig.~\ref{fig:sim_result} shows the energy spectra of particles emerging after neutrino propagation through 12~km of standard rock with a constant density of 2.65~$\rm{g/cm}^3$ when injecting neutrino fluxes with a flat spectrum $E^{-2}$  at 3-14~PeV assuming the equal flavor composition benchmark. For injected $\nu_\tau$ and $\bar{\nu}_\tau$, the spectra exhibit the characteristic features of tau regeneration at energies below the injected flux. As the thickness of a mountain is much shorter than the DIS mean free path (mfp), the attenuation of $\nu_\tau+\bar{\nu}_\tau$ is negligible. For injected $\bar{\nu}_e$, the effect of the Glashow resonance dominates, enhancing the interaction probability relative to DIS. Neutrinos and taus produced in the $W^-$ decay directly enter the propagation chain, generating neutrinos and taus at lower energies. In the end, taus from $\nu_\tau+\bar{\nu}_\tau$ and $\bar{\nu}_e$ that emerge after the propagation are at a comparable flux level in the Glashow resonance energy window. 

\begin{figure}[t]
    \centering
    \includegraphics[width=1\textwidth]{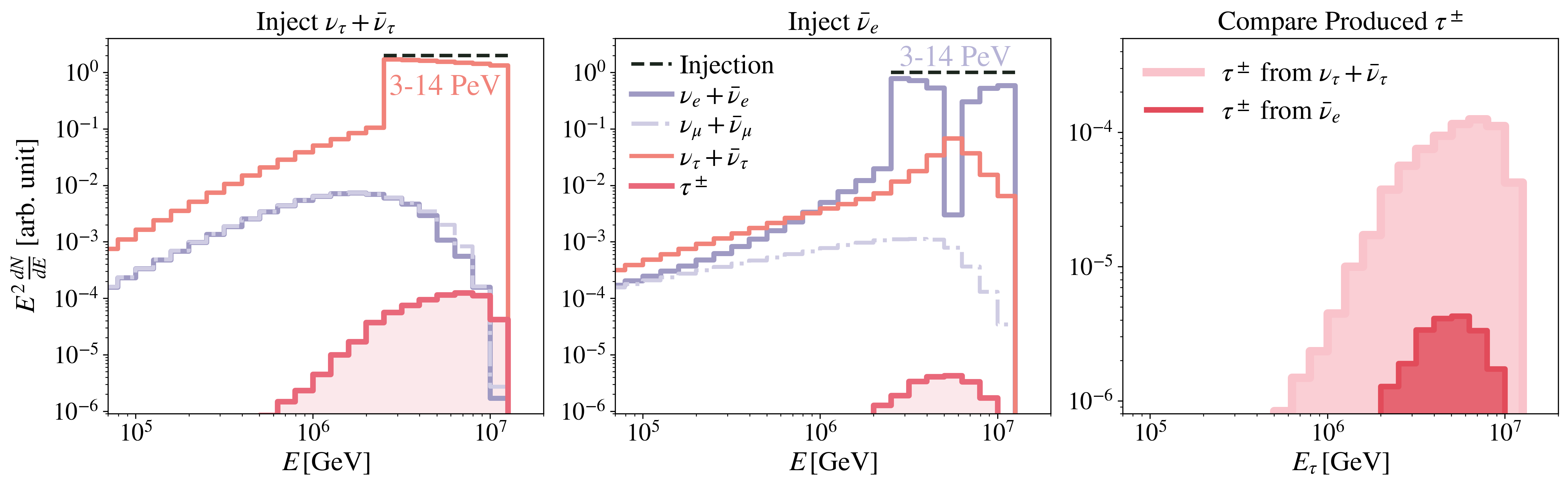}
    \caption{{\bf{{\textit{Energy Spectra at Surface for Earth-skimming neutrinos.}}}}
    Same as Fig.~\ref{fig:sim_result} but for Earth-skimming neutrinos arrived $4^\circ$ below the horizon.
    }
    \label{fig:sim_result_skimming}
\end{figure}

For the Earth-skimming case, Fig.~\ref{fig:sim_result_skimming} shows the case where the incidence angle is 4$^\circ$ below the horizon (889~km). As the distance in matter is significantly larger here compared to a mountain, the attenuation of incoming neutrinos due to DIS becomes more evident following the energy increase. Due to the short mfp of Glashow resonance, PeV $\bar{\nu}_e$ near the resonant energy experiences consequential attenuation where a large fraction of the flux is converted to $\tau$ lepton and neutrino fluxes, which leads to a competition between the $\bar{\nu}_e$ attenuation, and the production and survival of the taus afterwards. In the case of 4$^\circ$ below the horizon, the $\bar{\nu}_e$ generated tau flux level is about 2 orders lower than that generated by $\nu_{\tau}+\bar{\nu}_\tau$. For the Earth-skimming case, the distance in matter increases rapidly as the incident angle becomes larger.  We find that only when the incoming neutrinos are very close to the horizon, the contribution of $\bar{\nu}_e$ is non-negligible. Otherwise, the absorption of $\bar{\nu}_e$ dominates. While more taus can be produced, they decay fast and tau regeneration effect is less efficient at this energy and distance range combination. 

In Appendix~\ref{sec:appendix}, we show the spectra after propagation for both mountain-skimming and Earth-skimming cases that cover a mono-energetic injection and an injection with a broader energy range, as well as a smaller incident angle for Earth-skimming neutrinos in Fig.~\ref{fig:sim_result_3_14PeV_02deg}--\ref{fig:sim_result_mon_4deg}.

With the simulations, the acceptance of neutrino flavor $\alpha$ can be estimated by 
\begin{align}
\label{acc_eq}
    A^{\nu_\alpha}_{\rm{acc}}\left(E_\nu\right) = \int_{0}^{E_\nu} \int_{A_{\rm{geo}}} \int_{\Omega}  \frac{dP_{\rm{exit}}}{dE_\tau}(E_\nu,E_\tau)\cdot P^{\rm{visible}}_{{\rm{decay}}}\left(\,E_{\tau}\right)\cdot \epsilon_{{\rm{det}}}\left(E_{\tau}\right)dAd\Omega dE_\tau,
\end{align}
where $P_{{\rm{exit}}}$, $P^{\rm{visible}}_{{\rm{decay}}}$ and $\epsilon_{{\rm{det}}}$ are the probability of getting an exiting $\tau$ at $E_{\tau}$ generated from our simulation, the probability that the $\tau$ decays before reaching the detector to the visible channels and the detection efficiency of this event in the detector. Here, $P^{{visible}}_{\rm{decay}} = \mathrm{BR}^{{visible}} \cdot\left[ 1 - {\rm{exp}}\left( - D/\lambda^\tau_{\rm{decay}} \right)\right]$ where $\lambda^\tau_{\rm{decay}}(E_\tau)$ is the decay length of $\tau$ and the visible decay channels have a total branching ratio $\mathrm{BR}^{\rm{visible}} \simeq 82.6\%$, to account for $\tau$ decays to produce hadrons (64.8\%) or an electron (17.8\%). $D$ is the distance between the surface and the detector. $A_{\rm{geo}}$ is the physical area visible to the detector and $\Omega$ is the solid angle of the field of view. The computed acceptances are shown in Fig.~\ref{fig:acc}, which are also compared to the reported acceptances from TAMBO and TRINITY collaborations. 

The final event number in a telescope is 
\begin{equation}
    \frac{dN_\tau}{dE_\tau} (E_\tau) = T_{\rm{obs}}\int_{E_\nu}\Phi_{\nu_\alpha} \left(E_\nu\right)\cdot \frac{dA_{\text{acc}}^{\nu_\alpha}}{dE_\tau}(E_\nu, E_\tau) \,dE_\nu,
\end{equation}

where $\Phi_{\nu_\alpha}$ is the incoming flux of neutrino $\alpha$ and $T_{\rm{obs}}$ is the observing time of the detector including the duty cycle. Here, we do not include a smearing factor in our analysis to account for the energy reconstruction uncertainty.

\subsection{Mountain-skimming Signal}
\begin{figure}[t]
    \centering
    \includegraphics[width=1\textwidth]{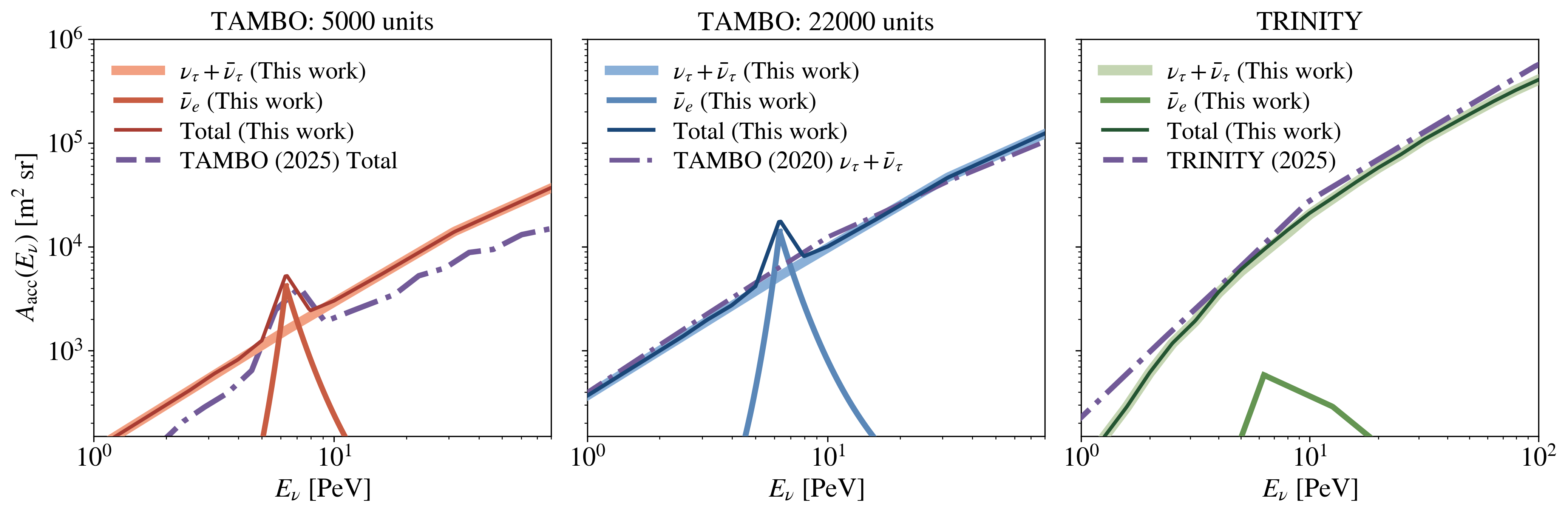}
    \caption{{\bf{{\textit{Neutrino acceptance for TAMBO and TRINITY.}}}} The detector acceptance is shown as a function of neutrino energy for TAMBO (left and middle) and TRINITY (right) configurations. Thick solid curves correspond to the combined $\nu_\tau + \bar{\nu}_\tau$ component, medium solid curves represent the $\bar{\nu}_e$ contribution arising from Glashow resonance interactions and the thin solid lines show their sum. In the TAMBO panels, both the original design (middle) and the updated configuration (left) are shown, and we compare our calculation with the acceptance reported in Refs. \cite{Romero-Wolf:2020pzh,TAMBO:2025jio} which are displayed as dash-dotted purple lines. In the right panel, we compare the acceptance we compute with the acceptance derived from its reported neutrino flux detection sensitivity~\cite{Stepanoff:2025vys}. 
  }
    \label{fig:acc}
\end{figure}

\begin{figure}[t]
    \centering
    \includegraphics[width=1\textwidth]{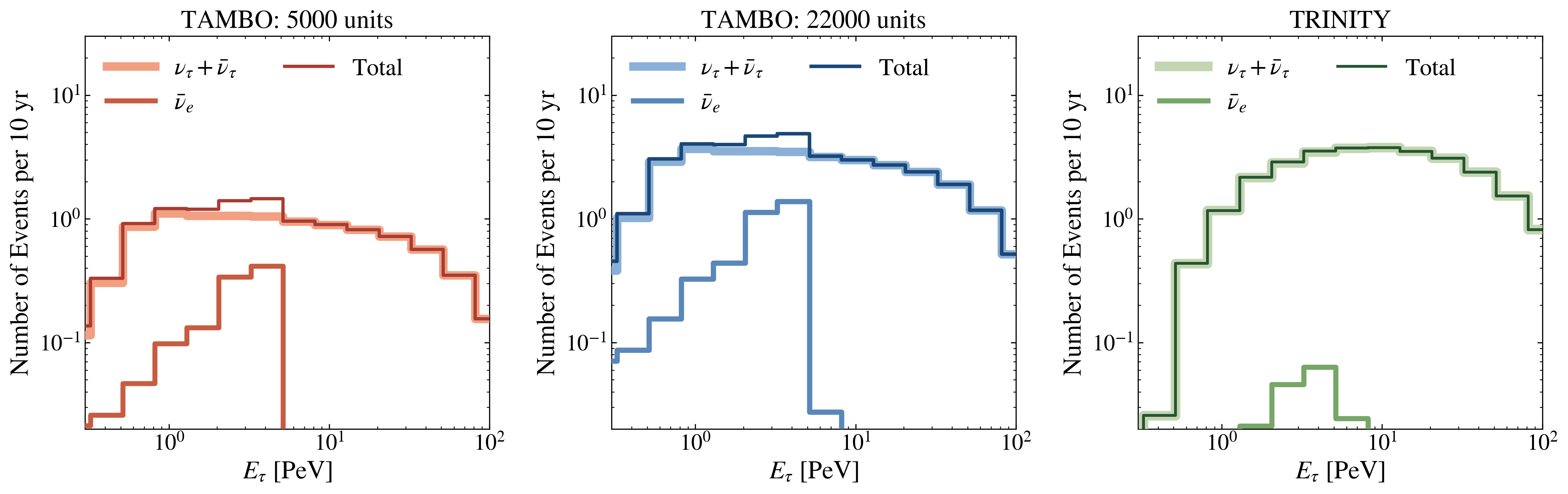}
    \caption{{\bf{{\textit{Expected tau event numbers with 10~yrs Operation.}}}}
    Tau event numbers for TAMBO 5000 units (left), TAMBO 22000 units (middle), and TRINITY (right), assuming an $E^{-2.5}$ spectrum consistent with the IceCube diffuse flux measurement~\cite{IceCube:2015gsk} and an equal flavor composition of all six flavors. 
    Thick curves denote the contribution from $\nu_\tau+\bar{\nu}_\tau$ via charged-current DIS, while medium curves represent the $\bar{\nu}_e$ contribution induced by the Glashow resonance. The thin lines correspond to the total event number. 
    }
    \label{fig:nof}
\end{figure}

We take TAMBO, which is a proposed deep-valley mountain-based observatory aiming at detecting mountain-skimming $\tau$ neutrinos in the 1--100~PeV energy range, as an example. In this work, we consider two configurations, matching the original design and the plan reported most recently. The detector layout remains the same, {\textit{i.e.}}, a 150~m spacing, while the number of detection units differs, corresponding to a full-size (22000 units) or a reduced-scale (5000 units) detector~\cite{Romero-Wolf:2020pzh,TAMBO:2025jio}. For each configuration, the geometric parameters used in our calculations, including the array layout, valley geometry, and effective injection surface, are implemented according to the corresponding proposal. Assuming a 150~m spacing of detection units and an average mountain slope of $\sim 35^\circ$, an effective geometric area would be $A_{\rm geo}= A_{\rm phys}\cos(35^\circ) \approx 100~{\rm{km}}^2$ for TAMBO with 22000 units, where $A_{\rm phys}$ denotes the total instrumented surface area determined by the array layout and spacing. The effective geometric area in the newer 5000 units configuration reduces to roughly $30\%$ of the original design.

Although various suitable sites for TAMBO exist including mountain valleys and fjords, Colca Valley in Peru is one strong possibility. With this location, we refer to the elevation map from the Shuttle Radar Topography Mission (SRTM) 1 arc-second global digital elevation model, accessed via the OpenTopography data portal, to trace a ray along each direction~\cite{3093, https://doi.org/10.1029/2005RG000183}. The field of view is discretized into angular bins in azimuth and elevation. For each direction, the neutrino trajectory is traced through the mountain profile to determine whether it traverses rock and emerges into the valley. From this geometric construction, we obtain the rock chord length along the trajectory and the exit point from the mountain. Only directions that satisfy the imposed geometric requirements, such as limits on the rock thickness and on the allowed exit region, are included in the acceptance calculation. The computed acceptance for TAMBO is shown in Fig.~\ref{fig:acc}. The acceptance of $\bar{\nu}_e$ exceeds that of tau neutrinos by a factor of about 3 around the resonant energy, and the total acceptance has a substantial increase. The reported acceptance by the TAMBO collaboration is shown for comparison. For the recent 5000-unit configuration, our total acceptance is slightly larger than TAMBO's reported acceptance; while for the previous 22000-unit, our $\nu_\tau+\bar{\nu}_\tau$ acceptance is consistent with that of TAMBO reported, showing that the inclusion of $\bar{\nu}_e$ in the propagation has negligible effect on the acceptance of tau neutrinos.

For a continuous operation over 10 years with a nominal duty cycle of 100\%, Fig.~\ref{fig:nof} demonstrates the expected number of events that can be observed by TAMBO's two configurations. Here we assume the neutrino flux of each flavor, including a separation of neutrinos and antineutrinos, contributes equally as a benchmark. The energy integrated event numbers in the resonance window are $\sim$6.7$f_{\bar{\nu}_e}$ (5000 units) and $\sim 22.4f_{\bar{\nu}_e}$ (22000 units) assuming a $E^{-2.5}$ spectrum, consistent with the IceCube combined diffuse flux measurement~\cite{IceCube:2015gsk}. $f_{\bar{\nu}_e}$ represents the fraction of ${\bar{\nu}_e}$. The neutrino spectrum is expected to affect the event number in this regime of low statistics. Varying the spectrum from the softest ($E^{-2.87}$) to the hardest ($E^{-2.37}$) reported by IceCube~\cite{Abbasi:2021qfz,IceCube:2020wum}, we find that the expected event number varies between  1.4$f_{\bar{\nu}_e}$ -- 7.4$f_{\bar{\nu}_e}$ for ${\bar{\nu}_e}$ in 5000-unit TAMBO and 4.6$f_{\bar{\nu}_e}$ -- 24.8$f_{\bar{\nu}_e}$ in 22000-unit TAMBO. 

\subsection{Earth-skimming Signal}
When very-high-energy and ultra-high-energy neutrinos travel nearly parallel to the Earth's surface below the horizon, they interact with the Earth matter to produce $\tau$. These neutrinos are referred to as Earth-skimming neutrinos and the $\tau$ can be detected through various techniques~\cite{ARA:2015wxq,GRAND:2018iaj,ANITA:2019wyx,ARIANNA:2019scz,PierreAuger:2019ens,Otte:2019knb,Prohira:2019glh,Wissel:2020sec,IceCube-Gen2:2020qha,PUEO:2020bnn,RNO-G:2020rmc,POEMMA:2020ykm}. While most of the telescopes have an energy threshold above tens of PeV, an air-shower imaging Cherenkov telescope (IACT) such as TRINITY can achieve a threshold down to PeV~\cite{Otte:2019knb}. The baseline design consists of 18 telescopes, each with a horizontal field of view of $60^\circ$ and a vertical field of view of $5^\circ$, providing full $360^\circ$ azimuthal coverage at an altitude of approximately $2$--$3~\mathrm{km}$ above ground. TRINITY is designed to detect air showers developing at distances up to $\sim 200~\mathrm{km}$ from the detector. The observable region is defined by the angular coverage of the telescopes in azimuth and elevation. For each emergence angle within this field of view, the neutrino propagation simulation utilizes the PREM Earth profile~\cite{Dziewonski:1981xy}. 

The geometric normalization of the acceptance is obtained by converting each solid-angle element of the telescope field of view into an effective surface element on the Earth. For an angular bin $\Delta\Omega$ corresponding to an emergence direction at distance $d$ from the detector, the associated surface patch is proportional to $\Delta\Omega d^{2}$. This relation follows from the geometric mapping between solid angle and area on a spherical surface. The resulting area element, together with the projection factor appropriate for Earth-skimming trajectories, provides the effective cross-sectional area through which neutrinos must pass to generate an observable event. The total acceptance is then obtained by summing over all angular bins the product of this geometric factor and the corresponding energy-dependent detection probability. Unlike TAMBO which operates as a continuously running surface array, IACTs are restricted to dark, clear nights. We take into account the 20\% duty cycle for TRINITY observations~\cite{Stepanoff:2025vys}. The right panel in Figure~\ref{fig:acc} shows the corresponding acceptance for the TRINITY configurations we compute compared to the acceptance derived from the diffuse neutrino flux sensitivity reported by the TRINITY collaboration~\cite{Stepanoff:2025vys} through
\begin{equation}
    A^{\nu_\tau+\bar{\nu}_\tau}_{{\rm{acc}}} = 3\cdot\frac{2.44 E_\nu}{T_{obs} \cdot {\rm{ln}}(10)\phi^{{tot}}_{{\rm{sens}}}(E_\nu)},
\end{equation}
where $\phi^{{tot}}_{{\rm{sens}}}$ is the all-flavor sensitivity flux. A factor 3 here corresponds to an equal flavor ratio assumption. Because TRINITY observes air showers developing at large distances from the detector, its effective geometric area grows approximately with the square of the observation distance, which leads to a rapidly increasing acceptance with energy for tau neutrinos. For $\nu_\tau+\bar{\nu}_\tau$, the acceptance we compute is consistent with that reported from TRINITY. The acceptance for $\bar{\nu}_e$ is substantially lower than that of $\nu_\tau+\bar{\nu}_\tau$. Around the resonance peak, the $\nu_\tau+\bar{\nu}_\tau$ acceptance is approximately 25 times larger than the $\bar{\nu}_e$ acceptance. This is because the Earth-skimming path length increases rapidly as the incident angle grows while mfp of the Glashow resonant interaction is short. Consequently, most primary \(\bar{\nu}_e\) interact well before reaching the exit point while more taus are produced. However, the decay length of tau is also short whereas the mfp of tau neutrino DIS is much longer at $\sim$PeV. Therefore, different from the mountain-skimming case, for most incoming Earth-skimming neutrinos, detecting the $\bar{\nu}_e$ contribution is inefficient. 

We show the expected number of events with 10 years of TRINITY exposure in Fig.~\ref{fig:nof}. The expected number of $\bar{\nu}_e$ induced events is orders of magnitude smaller, amounting to $\sim 1.0f_{\bar{\nu}_e}$ events over the same period, which contributes $\lesssim 1\%$ compared to $\nu_\tau+\bar{\nu}_\tau$ events for the nominal equal flavor ratio assumption and typical astrophysical neutrino production mechanisms. From the $E^{-2.87}$ to $E^{-2.37}$ spectrum assumption, the expected event number varies between 0.2$f_{\bar{\nu}_e}$  -- 1.2$f_{\bar{\nu}_e}$. Compared to TAMBO, detecting $\bar{\nu}_e$ as Earth-skimming events is more challenging. Although both experiments reach a sensitivity down to PeV, TRINITY is more sensitive at energies higher than 10~PeV with a larger expected tau neutrino event rate, which makes it more favored for ultra-high-energy tau neutrino detection.

\section{Sensitivity of the Neutrino Flavor Composition Measurement }\label{sec:flavor}

\subsection{Astrophysical Neutrino Flavor Composition}
\renewcommand{\arraystretch}{1.5}

\begin{table*}[t!]
    \centering
    \begin{tabularx}{\textwidth}{Z{1.7cm}Z{3.5cm}Z{6.5cm}Z{2.0cm}}
    \toprule
    \centering
    Production  & Source Flavor Ratio &  Earth Flavor Ratio & $\mathcal{R}=f_{\bar{\nu}_e}/f_{\nu_\tau+\bar{\nu}_\tau}$ \\
    \hline
    $pp$  & $\sixflavor{1}{1}{2}{2}{0}{0}$ &  $\sixflavor{0.17}{0.17}{0.16}{0.16}{0.16}{0.16}$ &  0.53 \\
    $pp\,\mu$ damped & $\sixflavor{0}{0}{1}{1}{0}{0}$ &  $\sixflavor{0.12}{0.12}{0.19}{0.19}{0.19}{0.19}$ & 0.32 \\
    $p\gamma$ & $\sixflavor{1}{0}{1}{1}{0}{0}$  &  $\sixflavor{0.26}{0.08}{0.20}{0.12}{0.20}{0.13}$ & 0.25 \\
    $p\gamma\,\mu$ damped &  $\sixflavor{0}{0}{1}{0}{0}{0}$ & $\sixflavor{0.24}{0.00}{0.37}{0.00}{0.39}{0.00}$ & 0.0 \\
    $n$ decay  & $\sixflavor{0}{1}{0}{0}{0}{0}$ &  $\sixflavor{0.00}{0.55}{0.00}{0.24}{0.00}{0.21}$ & 2.64 \\
    \bottomrule 
    \end{tabularx}
    \caption{{\bf{\textit{Flavor Compositions for Different Neutrino Production Models.}}} The flavor ratios of high-energy astrophysical neutrinos at the sources and at Earth arising from different production mechanisms. The flavor ratios are shown as \sixflavor{$f_{\nu_e}$}{$f_{ \bar{\nu}_e}$}{$f_{\nu_\mu}$}{$f_{\bar{\nu}_\mu}$}{$f_{{\nu}_\tau}$}{$f_{\bar{\nu}_\tau}$}. The standard neutrino oscillation effect is applied with the best-fit oscillation parameters from NuFIT6.1. The ratios of $\bar{\nu}_e$ to $\nu_\tau+\bar{\nu}_\tau$ fluxes are displayed in the last column.
    }
    \label{tab:flavor_ratio}
\end{table*}

High-energy cosmic rays produced in cosmic accelerators interact with gas or radiation in the vicinity, predominantly producing pions. High-energy astrophysical neutrinos are mainly produced from decays of these charged pions, {\textit{i.e.}} the \textit{pion decay} scenario, via $\pi^-\to \mu^- + \bar{\nu}_\mu$, followed by $\mu^-\to e^-+\bar{\nu}_e+\nu_\mu$, and their charge-conjugated processes. For the hadronuclear processes ($pp$), a uniform distribution of pion charges is expected, {\textit{i.e.}} $p+p\rightarrow n_\pi\left[\pi^-+\pi^++\pi^-\right]$ where $n_\pi$ is the multiplicity factor. For the photohadronic processes ($p\gamma$), the dominant interaction leading to neutrino production is $p+\gamma \rightarrow \Delta^+\rightarrow \pi^+ + n$ so that a majority $\pi^+$ contribution is expected.  The neutron may still escape the source and decay to produce neutrinos, but at much lower energies than neutrinos produced by the pion. The two scenarios cannot be distinguished in a 3-flavor composition measurement, since both of them lead to $(f_{e}:f_{\mu}:f_{\tau})_s = (1:2:0)$ at the sources. However, if we account for the asymmetry in charges and write neutrino and antineutrinos separately and write the 6-flavor composition as:$\sixflav{f_{\nu_e}}{f_{ \bar{\nu}_e}}{f_{\nu_\mu}}{f_{\bar{\nu}_\mu}}{f_{{\nu}_\tau}}{f_{\bar{\nu}_\tau}}$, there are $\sixflav{1}{1}{2}{2}{0}{0}$ for the $pp$ scenario and $\sixflav{1}{0}{1}{1}{0}{0}$ for the $p\gamma$ scenario, breaking the degeneracy. If a very strong magnetic field is present, the muons in the chain suffer from significant energy loss via synchrotron radiation and the decay would be consequently suppressed. This is the \textit{muon-damped} scenario and the ideal cases result in $\nu_\mu + \bar{\nu}_\mu$ production for $pp$ and $\nu_\mu$ production for $p\gamma$. In addition to the pion decay scenarios, there may be sources generating a pure $\bar{\nu}_e$ flux from a neutron beam~\cite{Anchordoqui:2003vc,Yasuda:2024fvc}. Protons and neutrons are produced from photodisintegration of Fe by photon fields. While protons can rapidly lose energy or get deflected by the magnetic field, the neutrons would decay, leading to a pure $\bar{\nu}_e$. This scenario is in general disfavored as the predominant contribution of neutrinos by the current 3-flavor measurement at TeV-PeV~\cite{Abbasi:2025fjc}. 

These neutrinos undergo flavor transitions en route from their sources to Earth. Because of the cosmological propagation distances and the uncertainty in the neutrino production locations, the transitions correspond to an averaged oscillation effect, with transition probability $P_{\alpha\rightarrow \beta} = \sum^3_i\left|U_{\alpha i}\right|^2\left|U_{\beta i}\right|^2$, where $U$ is the Pontecorvo–Maki–Nakagawa–Sakata (PMNS) matrix. Defining $\mathcal{R}=f_{\bar{\nu}_e}/{f_{\nu_\tau+\bar{\nu}_\tau}}$, and adopting the best-fit oscillation parameters from NuFIT6.1~\cite{Esteban:2024eli,nufit6.1} with normal mass ordering, we summarize in Table~\ref{tab:flavor_ratio} the flavor compositions at the sources and at Earth, together with $\mathcal{R}$, for each production mechanism. These predicted ratios determine the expected event numbers from the two contributions. 

For the tau air-shower experiments considered here, the flavor composition can only be inferred statistically, since the $\tau$ contributions from the two components are indistinguishable event by event. In our projections, we use the results of Section~\ref{sec:signal} to generate Asimov data. The Poisson likelihood is

\begin{equation}
   \mathcal{L} (\mathbf{\Theta} |{\rm{data}}) =  e^{-N_{\tau}} \prod_i \frac{N^{N^{\rm{obs}}_{\tau,\,i}}_{\tau,\,i}}{N^{\rm{obs}}_{\tau,\,i}!},
 \label{eq:likelihood}  
\end{equation}   
where $\mathbf{\Theta}$ denotes the free parameters, $N_\tau$ is the total number of $\tau$ events and $i$ runs over all energy bins; the superscript ``$\rm{obs}$" distinguishes observed from expected event numbers.

We assume the all-flavor astrophysical neutrino flux follows a power-law spectrum 
\begin{equation}
 \Phi_{\nu_\alpha}\left(E_\nu\right) = \Phi_{\nu_\alpha,\,0} \left( E_\nu/100\rm{TeV} \right)^{-\gamma} \cdot 10^{-18}\,\rm{GeV^{-1} cm^{-2}s^{-1}sr^{-1}}, 
\end{equation}
where $\Phi_{\nu_\alpha,\,0}$ and $\gamma$ are the flux normalization of ${\nu_\alpha}$ and the spectral index. We set the parameters $\Theta = \left(\right.\Phi_{\nu_\tau+\bar{\nu}_\tau,\,0}, \,\mathcal{R}, \, \gamma \left.\right)$ where $\mathcal{R} = \Phi_{\bar{\nu}_e,\,0} / \Phi_{\nu_\tau+\bar{\nu}_\tau,\,0} $ -- the ratio between the two fluxes. We compute the confidence intervals using the Feldman-Cousins approach~\cite{Feldman:1997qc}. Since there is a hard physical floor $\mathcal{R}\geq 0$, Wilks' theorem breaks down near the boundary. At each $\mathcal{R}$, we profile out $\gamma$ and $\Phi_{\nu_\tau+\bar{\nu}_\tau,\,0}$ and evaluate the test statistic $\rm{TS} = -2[\ln\mathcal{L}(\mathcal{R}) - \ln\mathcal{L}{\rm{\max}}]$. Then we generate Poisson datasets to construct the confidence intervals by using the confidence level (C.L.) percentile of their TS distribution as the critical threshold $\rm{TS}_{\rm crit}$. The confidence interval is therefore the set of $\mathcal{R}$ satisfying $\rm{TS}_{\rm obs} < \rm{TS}_{\rm crit}$. 

Because such experiments cannot observe $\nu_e$, $\nu_\mu$ and $\bar{\nu}_\mu$, a flavor composition study incorporating all neutrino flavors would require either additional assumptions on the contribution of unobservable flavors or a combination with experiments sensitive to other flavors; we therefore report our results in terms of $\mathcal{R}$ rather than as a flavor ternary or the 4-flavor representation discussed in Ref.~\cite{Liu:2023lxz}.

\subsection{Results}
\label{sec:results}
\begin{figure}[t]
    \centering
    \includegraphics[width=1\textwidth]{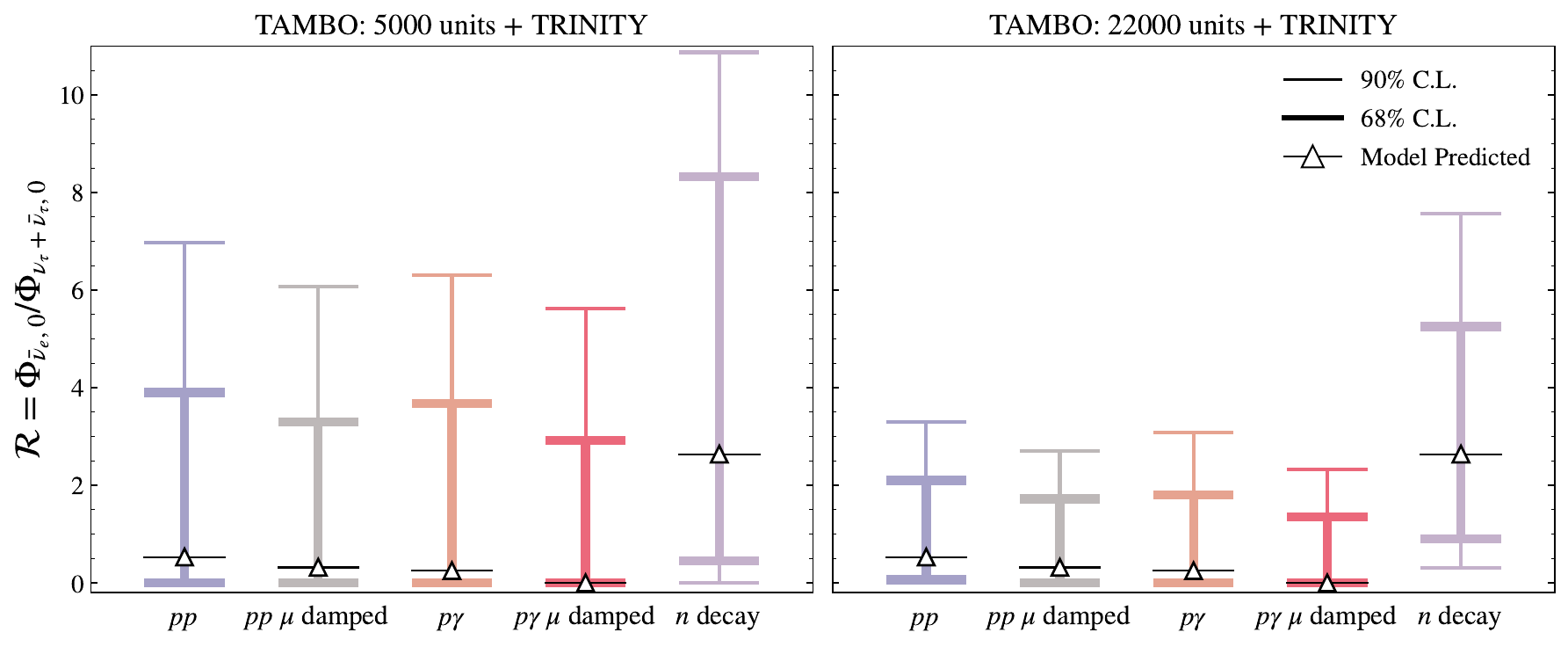}
    \caption{
    \textbf{\textit{Projection of the Flavor Ratio $\mathcal{R}$ for Benchmark Production Models.}} The projected measurement of $\mathcal{R}=\Phi_{\bar{\nu}_e,0}/\Phi_{\nu_\tau+\bar{\nu}_\tau,0}$ for different tested source production scenarios. The left and right panels correspond to 10~yr combined exposure of TAMBO 5000 units or TAMBO 22000 units and TRINITY, respectively. The thin and thick vertical bars denote the $90\%$ and $68\%$ C.L. intervals for each model.
    }
    \label{fig:result01}
\end{figure}

\begin{figure}[t]
    \centering
    \includegraphics[width=1\textwidth]{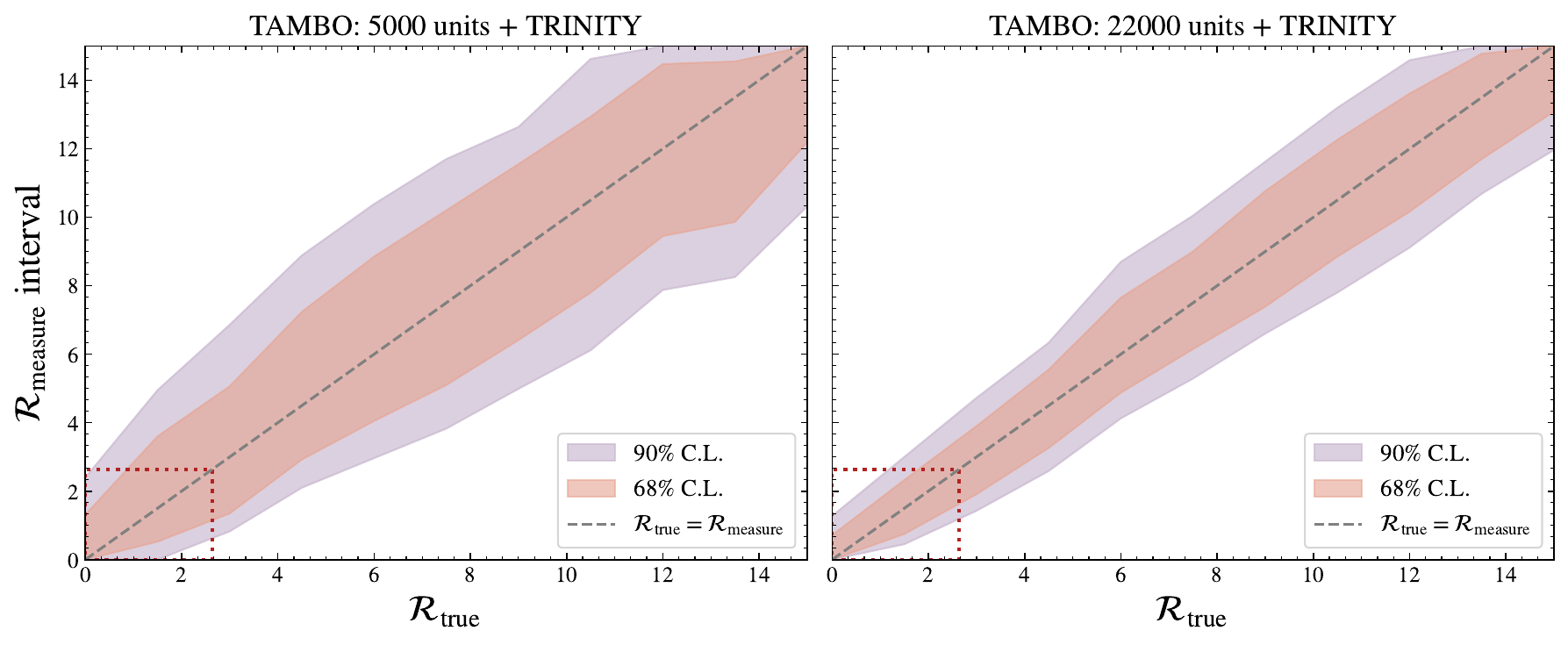}
    \caption{
    \textbf{\textit{Projection of the Flavor Ratio Measurement for Varying $\mathcal{R}$.}} 68\% and 90\% C.L. sensitivity intervals as a function of $\mathcal{R}$ without specifying a model. The small $\mathcal{R}$ region which corresponds to astrophysical neutrino production mechanisms with the standard neutrino oscillations is highlighted while a large $\mathcal{R}$ may arise due to new physics effects.   
    }
    \label{fig:result_arb_R}
\end{figure}

We evaluate the sensitivities for typical astrophysical production processes by combining the exposure of TAMBO and TRINITY. Here, we consider the two configurations of TAMBO. For the injection, we utilize the total flux normalization and $E^{-{2.5}}$ spectrum reported in Ref.~\cite{IceCube:2015gsk} with each flavor weighted by the ratio predicted by the model. The 68\% and 90\% C.L. sensitivity intervals are shown in Fig.~\ref{fig:result01} with values listed in Table~\ref{tab:tambo_ratio}. The $pp$ and $p\gamma$ scenarios, including their muon-damped cases, have intervals very close to each other for both C.L. and also the two configurations of TAMBO, indicating that differentiating these scenarios with a standalone $\mathcal{R}$ measurement in tau air-shower experiments is challenging. With a combined TAMBO 5000 units and TRINITY exposure, these scenarios can constrain $\mathcal{R}\lesssim 4\,(7)$ at $68\%(90\%)$ C.L. If the neutron decay scenario represents the true case, a wider interval is expected and the $pp$/$p\gamma$ scenarios can be excluded at $\sim$68\% C.L. When considering TAMBO 22000 units, the larger configuration substantially improves the measurement through increased statistics. Although discrimination remains challenging, the $\mathcal{R}$ measurement is able to constrain $\mathcal{R}\lesssim 2\,(3)$ at $68\%(90\%)$ C.L. for $pp$, $p\gamma$ and their muon-damped scenarios, which excludes the neutron decay model at 68\% C.L.; conversely, if the neutron decay case is true, $pp$ and $p\gamma$ models can be excluded at $\sim$90\% C.L. We note that the benchmark $p\gamma$ composition assumes a pure $\pi^+$ production from the $\Delta^+$ resonance. As the injected proton energy grows, multi-pion and other processes become non-negligible~\cite{Mucke:1999yb}; a $\pi^-$ component will thus appear and the pion-charge ratio shifts toward the $pp$ value, narrowing the separation between the two scenarios in $\mathcal{R}$.

We also compute the sensitivity intervals for different $\mathcal{R}$ covering a wider range without specifying a model, which are shown in Fig.~\ref{fig:result_arb_R}. Here, we keep the injected $\Phi_{\bar{\nu}_e}$ + $\Phi_{\nu_\tau + \bar{\nu}_\tau}$ flux the same as the total flux for all $\mathcal{R}$. When considering general astrophysical neutrino production processes with the standard neutrino oscillation effects, one expects the flavor composition sits in the central region of a flavor ternary plot, which corresponds to a small value of $\mathcal{R} \leq 2.64$. For new physics scenarios that introduce rich electron neutrino signal, a larger $\mathcal{R}$ may be expected which such experiments can readily identify (or exclude) via the the $\mathcal{R}$ measurement.

\section{Discussion \& Conclusion}
\label{sec:conclusion}
High-energy astrophysical neutrinos provide a unique probe of cosmic-ray acceleration, hadronic interactions in astrophysical environments, and fundamental neutrino physics over cosmological baselines. In particular, the flavor composition of the diffuse neutrino flux carries information about the neutrino production mechanism at the source and the flavor transition during propagation. While most flavor studies combine neutrinos and antineutrinos of the same flavor, the Glashow resonance provides a distinct handle on the $\bar{\nu}_e$ component near the PeV scale. Tau air-shower neutrino telescopes with PeV-scale thresholds therefore offer a complementary opportunity to probe the relative contributions of $\bar{\nu}_e$ and $\nu_\tau+\bar{\nu}_\tau$ through the observable sample of emerging $\tau$ leptons.

In this work, we incorporate the Glashow resonance into the neutrino and tau propagation calculation and evaluate the detection prospects in proposed mountain-skimming and Earth-skimming tau air-shower detectors. Mountain-skimming geometries yield substantially higher $\bar{\nu}_e$ acceptance than Earth-skimming ones due to shorter interaction distances comparable to the mfp of Glashow resonance while Earth-skimming $\bar{\nu}_e$ suffer from remarkable attenuation followed by quick decays of taus and inefficient tau regeneration. We estimate the sensitivity of measuring $\mathcal{R}$ which is the flavor ratio of $\bar{\nu}_e$ and $\nu_\tau+ \bar{\nu}_\tau$. Because the expected values of $\mathcal{R}$ are all small and close to each other, it is challenging to distinguish $pp$, $p\gamma$, $pp$ muon-damped and $p\gamma$ muon-damped scenarios with a standalone measurement in the currently designed configurations of TAMBO and TRINITY. However, we can optimistically expect an upper limit of $\mathcal{R}\lesssim$2 at 1$\sigma$ to be obtained and the $\bar{\nu}_e$-rich neutron decay model can be discriminated from them. Similarly, new physics models which introduce a large fraction of $\bar{\nu}_e$ may be well constrained even from a standalone measurement of $\mathcal{R}$. What to notice here is that the quantitative conclusions also depend strongly on the assumed spectrum, given the low statistics in this energy range.

In the future, observations from high-energy to ultra-high-energy neutrinos with distinct techniques and flavor identification approaches may be combined to construct a more comprehensive picture of the flavor composition of the neutrino sky, including neutrinos vs antineutrinos. The flavor composition measurements of TeV-PeV neutrinos by next-generation ice/water Cherenkov experiments are expected to be significantly improved compared to current IceCube results with boosted combined exposure and better flavor identification, including the capability of differentiating $\nu$ and $\bar{\nu}$ with the detection of Glashow resonant events at PeV~\cite{Song:2020nfh,Liu:2023lxz,Liu:2023flr,Tian:2025ymh}. A combination of such measurements with the $\nu_\tau+\bar{\nu}_\tau$ and $\bar{\nu}_e$ detection at PeV with experiments like TAMBO and TRINITY can further enhance the sensitivity as singling out electron and tau neutrinos has been persistently challenging in high-energy neutrino telescopes. At energies below $\sim$100~TeV, measuring the inelasticity of DIS of neutrinos with $\nu_\mu$-induced starting tracks offers a venue to tell $\nu/\bar{\nu}$ as the difference of cross section between neutrinos and antineutrinos enlarges as the energy decreases where the valence quarks become more dominant in the scatterings~\cite{Skrzypek:2025tmg}. In the ultra-high-energy regime, there is no established method to differentiate neutrinos and antineutrinos while a 3-flavor composition measurement is still feasible in Askaryan neutrino telescopes where a gain of sensitivity can be foreseen by combining observations of Earth-skimming tau air-shower experiments~\cite{Testagrossa:2023ukh,Coleman:2024scd}.

\acknowledgments
The authors would like to thank Carlos Argüelles, Ali Kheirandish and Aaron Vincent for helpful discussions and valuable comments. QL is supported by the Canada First Research Excellence Fund and the Natural Sciences and Engineering Research Council of Canada through the Arthur B. McDonald Canadian Astroparticle Physics Research Institute.

\bibliographystyle{JHEP}
\bibliography{biblio.bib}

\appendix
\section{Appendix}\label{sec:appendix}

At the PeV energy range which we want to focus on in this work, for $\bar{\nu}_e$, the mfp depends mainly on the Glashow resonant cross section where in rock there is $\lambda_{\rm{GR}}\simeq 26~{\rm{km}}$ at the resonant energy, while for charged-current DIS, the mfp in rock for PeV neutrinos, regardless of the flavor, has $\lambda_{\rm{CC}}\simeq$ 3000 -- 9000$~{\rm{km}}$, comparable to the size of the Earth. The mountain-skimming and Earth-skimming neutrinos correspond to two scenarios depending on the target length. The thickness of a mountain is $\sim \mathcal{O}\left(10~{\rm{km}}\right)$ whereas the distance in Earth increases rapidly from $\sim 20~{\rm{km}}$ to $\sim 1000~{\rm{km}}$ from 0.1$^\circ$ to 5$^\circ$ below the horizon. For a short target length comparable to the mfp of $\lambda_{\rm{GR}}$, one would anticipate efficient Glashow resonant interactions before the primary $\bar{\nu}_e$ flux is attenuated where sizable taus produced survive to escape the surface and be detected. For a long target length, given the short mfp, the primary $\bar{\nu}_e$ flux is expected to be attenuated quickly in matter while more $\tau$ and $\bar{\nu}_\tau$ are produced relatively close to the incident location. Most taus decay before escaping the surface and the $\bar{\nu}_\tau$ does not have a propagation distance long enough to produce taus, making the tau regeneration effect inefficient.

\begin{figure}
    \centering
    \includegraphics[width=1\textwidth]{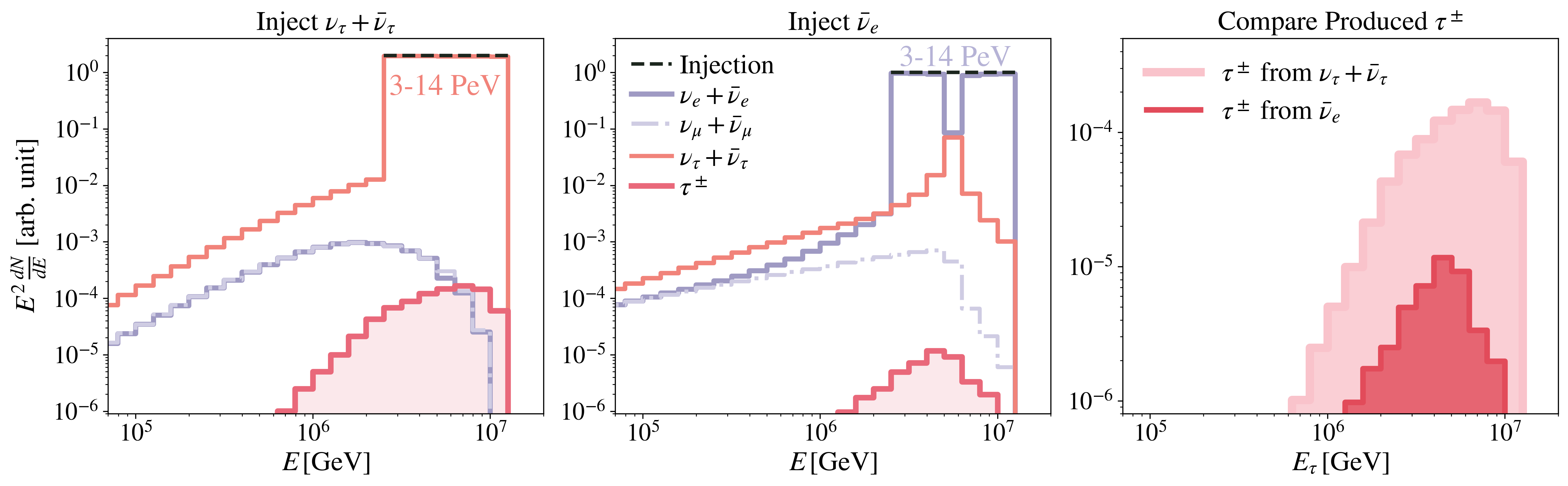}
    \caption{
    \textbf{\textit{Energy Spectra at Surface for Earth-skimming neutrinos.}} Same as Fig.~\ref{fig:sim_result_skimming} but for Earth-skimming neutrinos arrived $0.2^\circ$ (44~km) below the horizon.
    }
\label{fig:sim_result_3_14PeV_02deg}
\end{figure}

\begin{figure}
    \centering
    \includegraphics[width=1\textwidth]{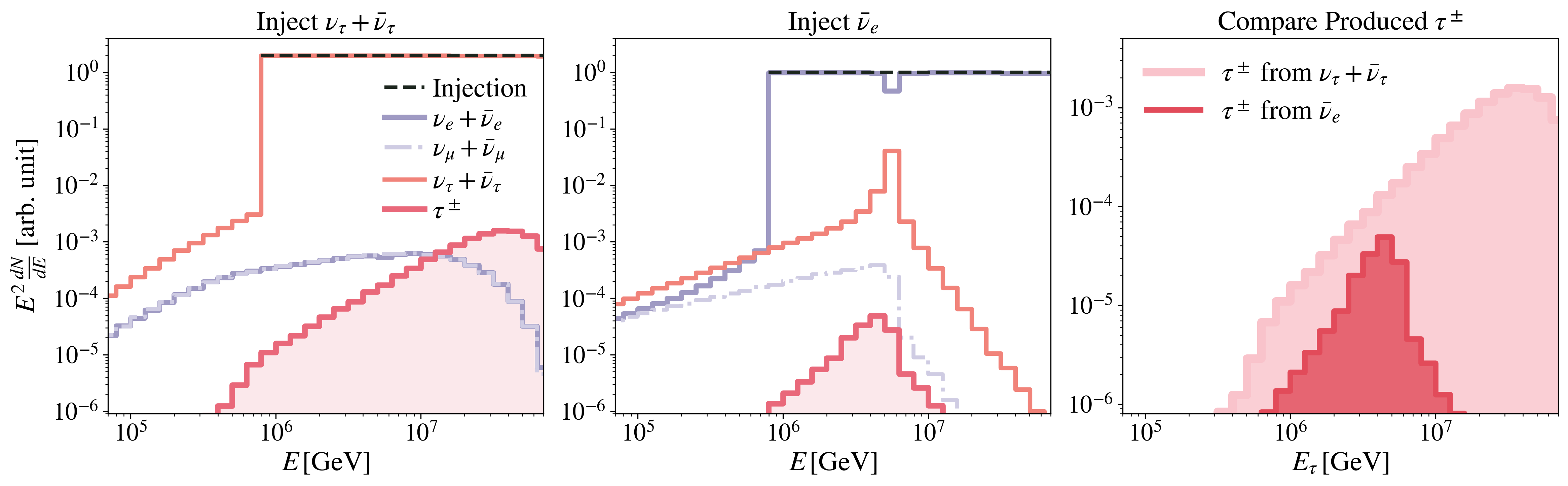}
    \caption{
    \textbf{\textit{Energy Spectra at Surface for Mountain-skimming Neutrinos with Broader Energy-Range Injection.}} Same as Fig.~\ref{fig:sim_result} for neutrino propagation through 12~km of standard rock but covers an injection at 1--100~PeV.
    }
\label{fig:sim_result_12}
\end{figure}

\begin{figure}
    \centering
    \includegraphics[width=1\textwidth]{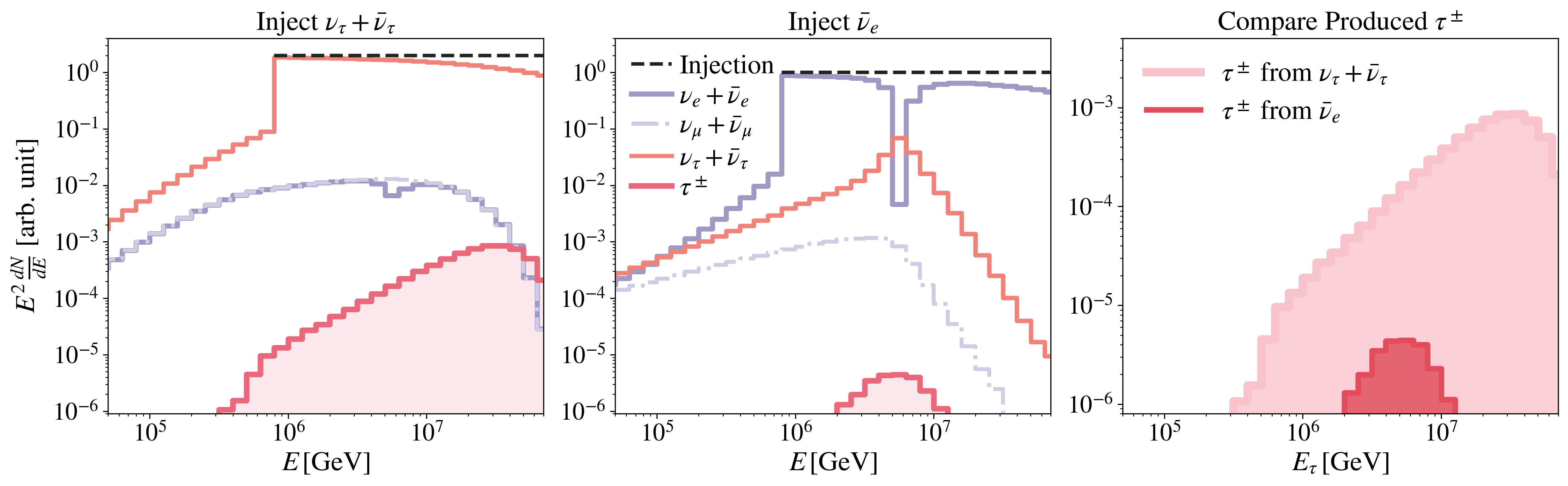}
    \caption{\textbf{\textit{Energy Spectra at Surface for Earth-skimming Neutrinos with Broader Energy-Range Injection.}} Same as Fig.~\ref{fig:sim_result_skimming} for Earth-skimming neutrinos arrived 4$^\circ$ below the horizon but covers an injection at 1--100~PeV.
    }
\label{fig:sim_result_444}
\end{figure}

\begin{figure}
    \centering
    \includegraphics[width=1\textwidth]{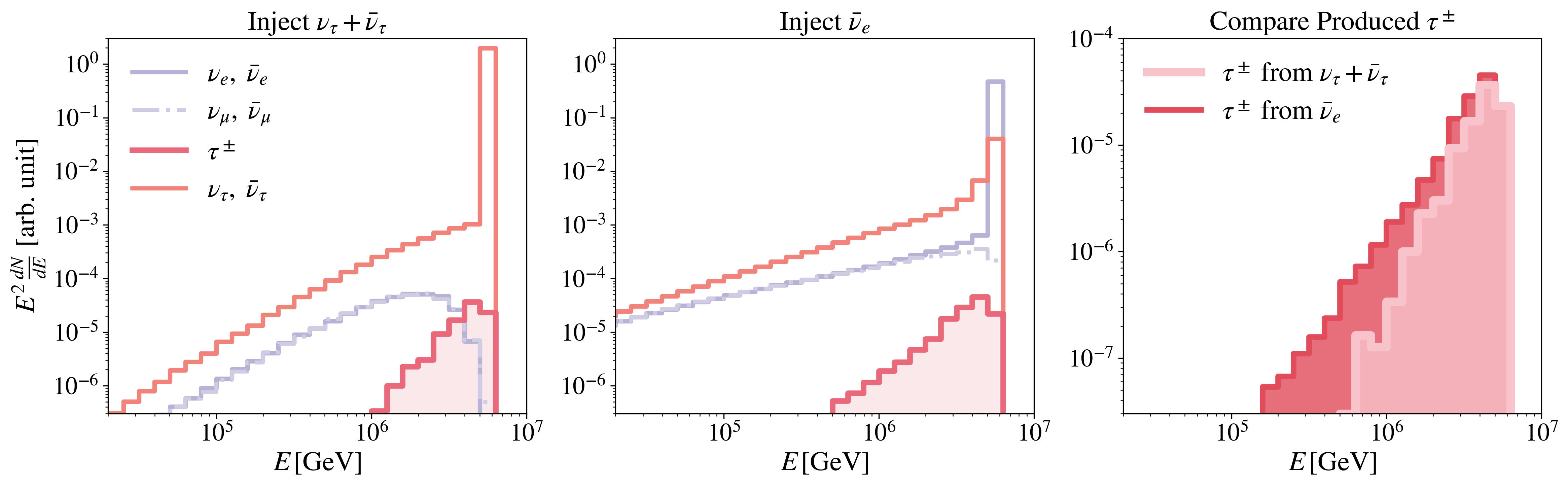}
    \caption{\textbf{\textit{Energy Spectra at Surface for Mono-energetic Mountain-skimming Neutrinos.}} Same as Fig.~\ref{fig:sim_result} but the injection is mono-energetic at $E_\nu=6.3$ PeV.
    }
\label{fig:sim_result_mono_12}
\end{figure}

\begin{figure}
    \centering
    \includegraphics[width=1\textwidth]{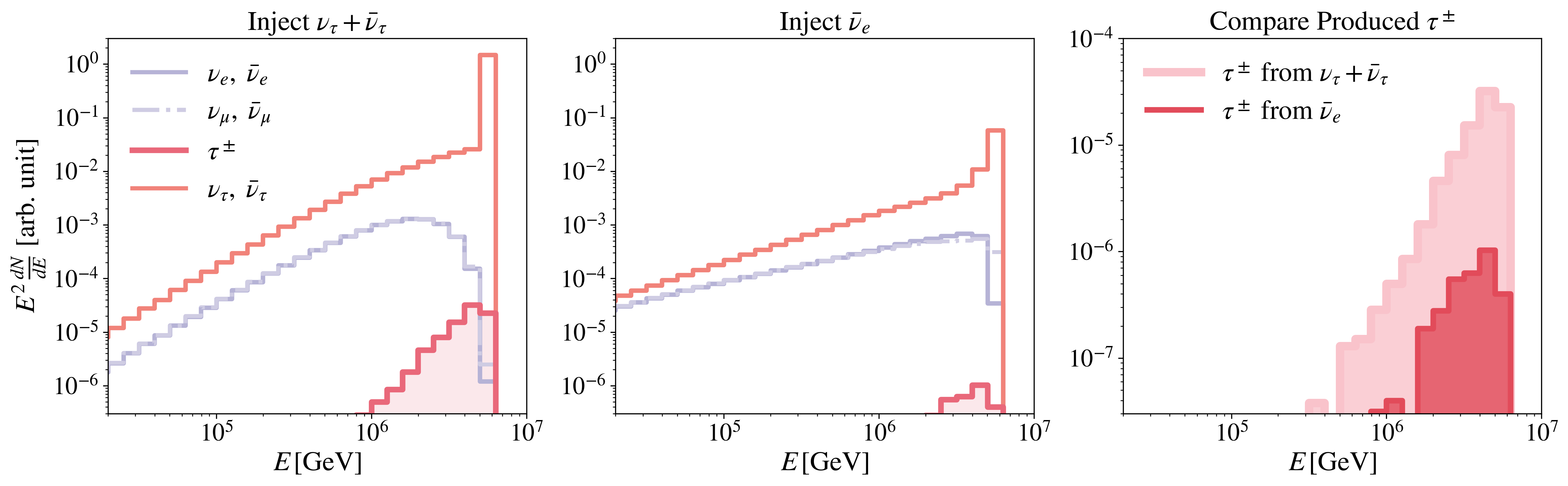}
    \caption{\textbf{\textit{Energy Spectra at Surface for Mono-energetic Earth-skimming Neutrinos.}} Same as Fig.~\ref{fig:sim_result_skimming} but the injection is mono-energetic at $E_\nu=6.3$ PeV.
    }
\label{fig:sim_result_mon_4deg}
\end{figure}

Figures~\ref{fig:sim_result_3_14PeV_02deg}--\ref{fig:sim_result_mon_4deg} show supplementary energy spectra of neutrinos and taus at the surface for both the mountain-skimming and Earth-skimming propagation cases. These figures extend the examples shown in Figs.~\ref{fig:sim_result} and~\ref{fig:sim_result_skimming} by considering a smaller incident angle, a broader injection over $1$--$100$ PeV and a mono-energetic injection at the Glashow-resonant energy. Comparing Figs.~\ref{fig:sim_result},~\ref{fig:sim_result_skimming} and~\ref{fig:sim_result_3_14PeV_02deg}, they show as the propagation length grows, the efficiency of tau production from $\bar{\nu}_e$ drops fast. When considering an extended energy for both detection cases, as shown in Figs.~\ref{fig:sim_result_12} and~\ref{fig:sim_result_444}, the tau flux generated from $\bar{\nu}_e$ does not change much while the flux generated from tau neutrinos is slightly larger at $\sim$PeV due to more efficient tau regeneration effects at higher energies. Figures ~\ref{fig:sim_result_mono_12} and ~\ref{fig:sim_result_mon_4deg} correspond to mono-energetic injections at the Glashow resonant energy for both the Mountain-skimming and Earth-skimming cases.


\begin{table*}
\centering 
\setlength{\tabcolsep}{6pt}
\begin{tabularx}{\textwidth}{Z{2.5cm}Z{1.8cm}Z{2.2cm}Z{2.2cm}Z{2.2cm}Z{2.2cm}}
\toprule
\multirow{2}{*}{Model} & \multirow{2}{*}{Predicted $\mathcal{R}$} & \multicolumn{2}{c}{TAMBO 5000 + TRINITY} & \multicolumn{2}{c}{TAMBO 22000 + TRINITY} \\
\cmidrule(lr){3-4} \cmidrule(lr){5-6}
& & 68\% C.L. & 90\% C.L. & 68\% C.L. & 90\% C.L. \\
\midrule
$pp$ & 0.53 & 0.0--3.9 & 0.0--7.0 & 0.1--2.1 & 0.0--3.3 \\
$pp$ $\mu$ damped & 0.32 & 0.0--3.3 & 0.0--6.1 & 0.0--1.7 & 0.0--2.7 \\
$p\gamma$ & 0.25 & 0.0--3.7 & 0.0--6.3 & 0.0--1.8 & 0.0--3.1 \\
$p\gamma$ $\mu$ damped & 0.00 & 0.0--2.9 & 0.0--5.6 & 0.0--1.3 & 0.0--2.3 \\
$n$ decay & 2.64 & 0.4--8.3 & 0.0--10.9 & 0.9--5.3 & 0.3--7.6 \\
\bottomrule
\end{tabularx}
\caption{\textbf{\textit{Confidence Intervals of the Flavor Ratio $\mathcal{R}$.}} Projected Feldman-Cousins 68\% and 90\% confidence intervals for different neutrino production scenarios considering combined observations of TAMBO 5000 units and TRINITY or TAMBO 22000 units and TRINITY. }
\label{tab:tambo_ratio}
\end{table*}

\end{document}